\documentclass[fleqn,usenatbib]{mnras}

\usepackage{newtxtext,newtxmath}

\usepackage[T1]{fontenc}

\usepackage{graphicx}	
\usepackage{amsmath}	
\usepackage{color}
\usepackage{soul}
\usepackage{hyperref}
\usepackage{xspace}
\usepackage{deluxetable}
\usepackage{afterpage}

\hypersetup{    
  colorlinks      = {true},
  linkcolor       = {blue},
  citecolor       = {blue},
  urlcolor        = {blue},
}

\newcommand{\code}[1]{\texttt{#1}\xspace}

\newcommand{\gaia}{\textit{Gaia}\xspace}
\newcommand{\Gaia}{\gaia}

\newcommand{\unit}[1]{\ensuremath{\mathrm{\,#1}}\xspace}

\newcommand{\Teff}{\ensuremath{T_\mathrm{eff}}\xspace}
\newcommand{\logg}{\ensuremath{\log\,g}\xspace}

\newcommand{\kms}{\unit{km\,s^{-1}}}

\title[Typhon Chemical Abundances]{Chemical Abundances of the Typhon Stellar Stream\thanks{This paper includes data gathered with the 6.5~meter Magellan Telescopes located at Las Campanas Observatory, Chile.}}

\author[A. P. Ji et al.]{
Alexander P. Ji,$^{1,2}$\thanks{E-mail: alexji@uchicago.edu}
Rohan P. Naidu,$^{3}$
Kaley Brauer,$^{4}$
Yuan-Sen Ting,$^{5,6}$
and Joshua D. Simon$^{7}$
\\
$^{1}$ Department of Astronomy \& Astrophysics, University of Chicago, 5640 S Ellis Avenue, Chicago, IL 60637, USA\\
$^{2}$ Kavli Institute for Cosmological Physics, University of Chicago, Chicago, IL 60637, USA\\
$^{3}$ Center for Astrophysics | Harvard \& Smithsonian, 60 Garden Street, Cambridge, MA 02138, USA\\
$^{4}$ {Department of Physics and Kavli Institute for Astrophysics and Space Research, Massachusetts Institute of Technology, Cambridge, MA 02139, USA}\\
$^{5}$ {Research School of Astronomy \& Astrophysics, Australian National University, Cotter Rd., Weston, ACT}\\
$^{6}$ {School of Computing, Australian National University, Acton, ACT 2601, Australia}\\
$^{7}$ {Observatories of the Carnegie Institution for Science, 813 Santa Barbara Street, Pasadena, CA 91101, USA}
}

\date{Accepted XXX. Received YYY; in original form ZZZ}

\pubyear{2022}

\begin{document}
\label{firstpage}
\pagerange{\pageref{firstpage}--\pageref{lastpage}}
\maketitle

\begin{abstract}
We present the first high-resolution chemical abundances of seven stars in the recently discovered high-energy stream Typhon. Typhon stars have apocenters ${\gtrsim}100$ kpc, making this the first detailed chemical picture of the Milky Way's very distant stellar halo.
Though the sample size is limited, we find that Typhon's chemical abundances are more like a dwarf galaxy than a globular cluster, showing a metallicity dispersion and no presence of multiple stellar populations.
Typhon stars display enhanced $\alpha$-element abundances and increasing $r$-process abundances with increasing metallicity.
The high-$\alpha$ abundances suggest a short star formation duration for Typhon, but this is at odds with expectations for the distant Milky Way halo and the presence of delayed $r$-process enrichment.
If the progenitor of Typhon is indeed a new dwarf galaxy, possible scenarios explaining this apparent contradiction include a dynamical interaction that increases Typhon's orbital energy, a burst of enhanced late-time star formation that raises [$\alpha$/Fe], and/or group preprocessing by another dwarf galaxy before infall into the Milky Way.
Alternatively, Typhon could be the high-energy tail of a more massive disrupted dwarf galaxy that lost energy through dynamical friction.
We cannot clearly identify a known low-energy progenitor of Typhon in the Milky Way, but 70\% of high-apocenter stars in cosmological simulations are from high-energy tails of large dwarf galaxies.
Typhon's surprising combination of kinematics and chemistry thus underscores the need to fully characterize the dynamical history and detailed abundances of known substructures before identifying the origin of new substructures.
\end{abstract}




\section{Introduction}

The Milky Way galaxy's stellar halo is expected to be full of stellar streams from accreted dwarf galaxies \citep[e.g.][]{Bullock2005,Helmi2020}.
A few dwarf galaxy streams were discovered in early surveys of the Milky Way's stellar halo \citep{Helmi1999,Majewski2003}, and the number of stellar streams grew rapidly \citep{Belokurov2006, Grillmair2009, Bernard2016, Shipp2018} with the advent of deep photometric surveys like the Sloan Digital Sky Survey \citep{York2000}, Pan-STARRS \citep{Chambers2016}, and the Dark Energy Survey, \citep{DarkEnergySurveyCollaboration2016} though many of these are from tidally disrupting globular clusters. 
The number of streams and their full 6D dynamical characterization has recently exploded \citep{Price-Whelan2018,Shipp2019,Ibata2019,Ibata2021,Li2019,Li2022,Mateu2022} due to precise and all-sky astrometry from the \Gaia satellite \citep{GaiaCollaboration2016,Lindegren2021,Katz2022,GaiaCollaboration2022DR3}.
Due to \Gaia's relatively bright magnitude limit, most of the stellar streams are very close to the Sun, though streams initially identified just with deep photometry tend to be found at larger distances \citep{Bonaca2021,Li2022,Martin2022,Malhan2022}.

Recently, a new candidate dwarf galaxy stream on a high-energy polar orbit named Typhon was discovered in the \gaia DR3 radial velocity sample \citep{Tenachi2022, Dodd2022}. Typhon stars have apocenters \emph{above 100 kpc}, some of the most distant apocenters known in the Galaxy.
\citet{Tenachi2022} used metallicities of seven Typhon stars in LAMOST \citep{Wang2022} to show it is likely a dwarf galaxy, though the result depended primarily on a single metal-poor star in the system.
Other than the Sgr stream, most known dwarf galaxy streams have apocenters ${\lesssim}60$\,kpc, so Typhon may be providing one of the first glimpses of the disrupted dwarf galaxy population of the outer halo, filling in a current dearth of dwarf galaxy streams with large apocenters that are expected in an unbiased stream population \citep{Li2022}.
Despite its large apocenter, Typhon is also one of the \emph{closest} dwarf galaxy stream candidates discovered to date with a galactocentric pericenter of about 6\,kpc and all known member stars within 5\,kpc of the sun \citep{Tenachi2022,Dodd2022}.
Thus, its stars are easily amenable to high-resolution spectroscopic followup, in contrast to direct chemical studies of the outer halo that require a very large investment of observing time \citep[e.g.][]{Battaglia2017,Hill2019}.
In general, theoretical expectations for outer halo dwarf galaxies is that they should have accreted at relatively late times to retain large apocenters \citep[e.g.][]{Rocha2012} and thus have low $\alpha$-element abundances because their star formation was not quenched until recently \citep[e.g.][]{Johnston2008}.

Here we present the first chemical abundances derived from high-resolution spectroscopy of the Typhon stream.
Our goals are to confirm that Typhon's progenitor is a dwarf galaxy and not a globular cluster, to study the chemical evolution of Typhon, and to understand its physical origin.
Section~\ref{sec:methods} describes the observations and abundance analysis.
Section~\ref{sec:results} shows the chemical abundances of Typhon compared to globular clusters and dwarf galaxies, where we find Typhon is relatively $\alpha$-rich.
We discuss formation scenarios for Typhon and conclude in Section~\ref{sec:conclusion}.

\begin{table*}
    \centering
    \setlength{\tabcolsep}{3pt}
    \begin{tabular}{l|ccc|cc|cc|cc|cc|cc}
source\_id & RA & Dec & $v_{\rm hel}$ & SNR & SNR & $\Teff$ & $\sigma_{\Teff}$ & $\logg$ & $\sigma_{\logg}$ & $\nu_t$ & $\sigma_{\nu t}$ & [M/H] & $\sigma_{\mbox{[M/H]}}$ \\
& h:m:s & d:m:s & \kms & 4500{\AA} & 6500{\AA} & K & K & cgs & cgs & \kms & \kms & & \\ \hline
3913243629368310912 & 11:53:15.8 & $+$10:37:44.5 & $+301.4$ &  79 & 140 & 4710 & 112 & 1.50 & 0.22 & 1.88 & 0.26 & $-1.49$ & 0.22 \\
3891712266823336192 & 12:07:27.4 & $+$01:40:48.0 & $+322.1$ &  71 & 121 & 4725 & 114 & 1.25 & 0.21 & 1.77 & 0.25 & $-1.51$ & 0.22 \\
1765600930139450752 & 21:49:48.6 & $+$10:48:42.9 & $-247.5$ & 105 & 184 & 4755 & 102 & 1.55 & 0.20 & 1.45 & 0.21 & $-2.25$ & 0.20 \\
3736372993468775424 & 13:11:51.4 & $+$11:17:22.0 & $+378.0$ &  70 & 110 & 4835 & 110 & 1.85 & 0.21 & 1.50 & 0.24 & $-1.65$ & 0.22 \\
3793377208170393984 & 11:33:56.8 & $-$02:28:21.0 & $+300.6$ &  58 &  86 & 5355 & 115 & 1.80 & 0.21 & 2.34 & 0.24 & $-1.70$ & 0.21 \\
3573787693673899520 & 12:04:03.9 & $-$13:22:10.4 & $+395.5$ &  45 &  62 & 5810 & 121 & 3.45 & 0.22 & 1.47 & 0.24 & $-1.57$ & 0.22 \\
3939346894405032576 & 13:20:03.0 & $+$19:41:41.9 & $+374.9$ &  34 &  46 & 6050 & 167 & 3.70 & 0.22 & 1.55 & 0.28 & $-1.69$ & 0.24 \\
    \end{tabular}
    \caption{Observations and stellar parameters for seven stars in Typhon, sorted by effective temperature.
    The radial velocity $v_{\rm hel}$ is from our MIKE data.
    SNR is signal-to-noise ratio per pixel.
    [M/H] is the model atmosphere metallicity in our analysis, which can differ slightly from [Fe/H].
    }
    \label{tab:obs}
\end{table*}

\section{Observations and Abundance Analysis}\label{sec:methods}

\begin{figure*}
    \centering
    \includegraphics[width=\linewidth]{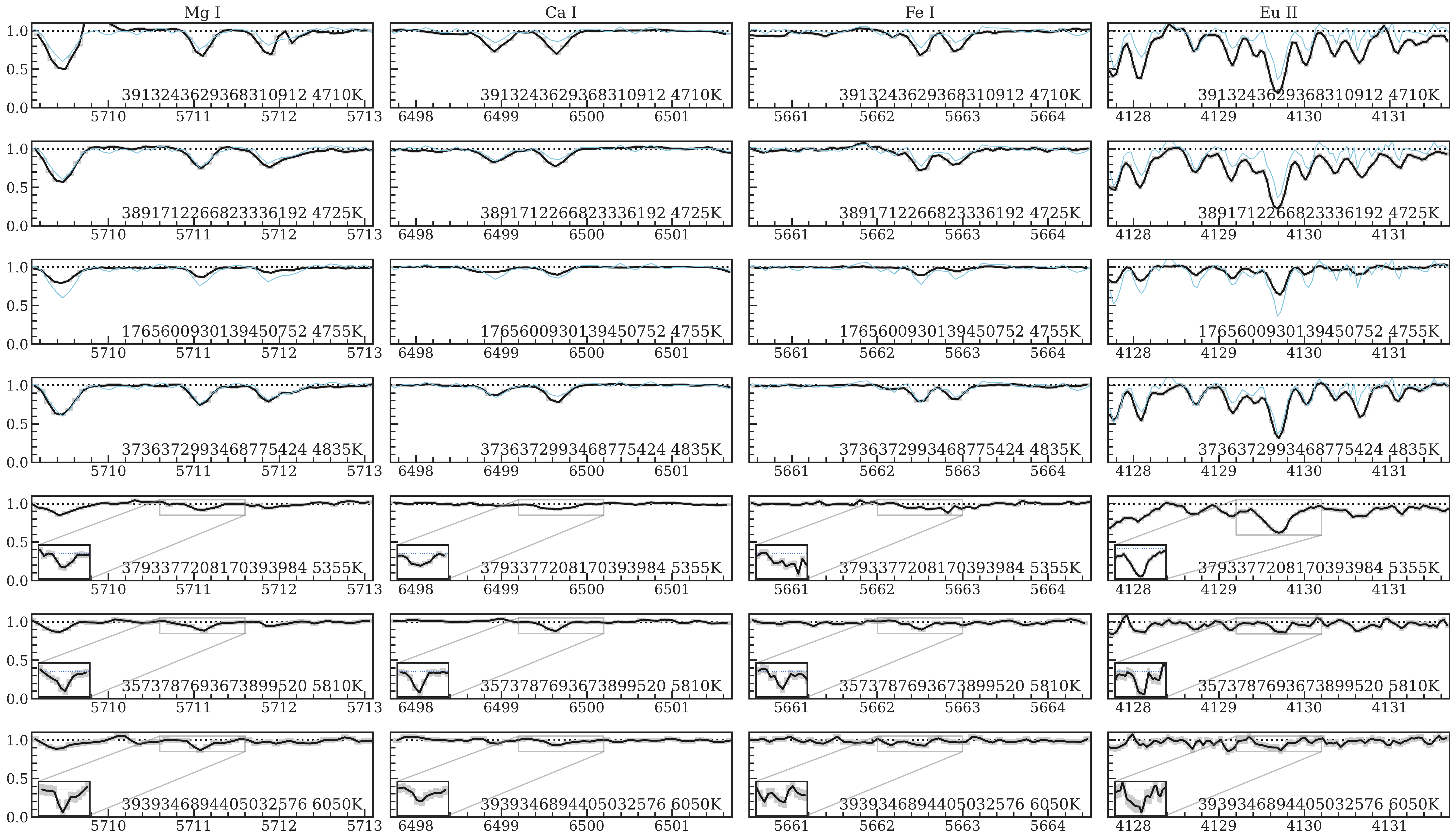}
    \caption{
    Spectra for Typhon stars, sorted by $\Teff$ (cool to hot from top to bottom) centered around selected absorption lines.
    Hotter or rotating stars have weaker lines, so a zoomed-in inset is shown.
    For comparison, the star Jhelum\_2\_2 \citep{Ji2020b} is plotted in cyan in the top four panels, which has similar temperature and metallicity as star 3736372993468775424.
    Note the excess of calcium and europium in the Typhon star compared to the Jhelum star, while Mg and Fe have more similar strength.
    Star 3793377208170393984 is a RHB star with a high rotational velocity causing its lines to be very broadened.
    }
    \label{fig:spectra}
\end{figure*}

We selected all seven Typhon member stars from \citet{Tenachi2022} that could be observed from Las Campanas Observatory during a recent observing run.
We will refer to these stars by their full \Gaia DR3 source IDs, but the reader can track the stars through the paper with the first four digits of the source ID, which are unique across the seven target stars.
High-resolution spectra were obtained on 28 Jun 2022 with the 0\farcs7 slit on Magellan/MIKE \citep{Bernstein2003}. The exposure time was 20 minutes per star.
Data were reduced using MIKE reduction pipeline\footnote{\url{https://code.obs.carnegiescience.edu/mike}} in the CarPy software package \citep{Kelson2003}. Radial velocities and heliocentric corrections were found using \code{smhr}\footnote{\url{https://github.com/andycasey/smhr}} \citep{Casey2014}.
The stars and radial velocities are given in Table~\ref{tab:obs}. 
The formal cross correlation radial velocity uncertainties are 0.1-0.2 \kms, but past experience has shown that slit centering and wavelength calibration sets a ${\approx}1\kms$ uncertainty floor that we have not attempted to address here \citep{Ji2020a}.
Examples of the spectra are shown in Figure~\ref{fig:spectra}.

Comparing the observed velocities to the \Gaia DR3 radial velocity of $-271.9$ \kms, we found star 1765600930139450752 to clearly have velocity variability, differing by 24.4 \kms compared to the MIKE velocity of $-247.5\kms$.
This is rather suspicious, as star 1765600930139450752 is the most metal-poor star in Typhon by a wide margin, and we will rely heavily on the membership of this star in Typhon for our interpretation.
However, the velocity difference is not large enough to affect this star's membership in Typhon.
The \Gaia DR3 \code{nss\_two\_body\_orbit} table gives a binary solution using 24 observations for a period of $73.98 \pm 0.16$\, days, center of mass velocity of $-252.3 \pm 0.6$ \kms, and semi-amplitude $23.4 \pm 1.3$ \kms.
Recomputing the orbits with this center of mass (CM) velocity in \code{gala} \citep{gala,adrian_price_whelan_2020_4159870} did not result in a substantially different orbit compared to the Gaia DR3 radial velocity (as expected, since the total CM velocity of $493.0 \kms$ changes by $<$5\%).
The CM apocenter and pericenter are 163 kpc and 4.9 kpc, compared to 173 kpc and 4.9 kpc when using the DR3 velocity.
Thus, we concur with \citet{Tenachi2022} that there is no reason to doubt that this star is part of Typhon.

We normalized and stitched the spectral orders and measured equivalent widths with \code{smhr}.
To speed up the normalization and equivalent width measurement, we used \code{Payne4MIKE} (\citealt{Ting2019}, Ji et al., in prep), which does a full spectrum fit to the MIKE data. \code{Payne4MIKE} eventually aims to infer $\Teff$, $\logg$, [M/H], [$\alpha$/Fe], and rotational velocity; but we did not use the fitted stellar parameters as they are not yet calibrated. However, the best-fitting \code{Payne4MIKE} spectrum is still very useful for identifying where absorption lines should occur. We thus used the best-fit synthetic spectrum to initialize \code{smhr} files masking significant absorption lines.
For equivalent widths, we adopted the line list from \citet{Ji2020b}, which was used to analyze seven other stellar streams. Except for a few specific elements, we removed all lines with $\lambda < 4500$\,{\AA}, as the Typhon stars are all relatively metal-rich compared to the stars in \citet{Ji2020b}, so for most species there was no need to use lines below 4500{\AA} where increased blending, more difficult continuum placement, and lower S/N made accurate analysis more difficult.
Equivalent widths of ${\approx}300$ lines were fit with Gaussian profiles and carefully inspected in \code{smhr}. Lines requiring Voigt profile fits were rejected.

Spectroscopic stellar parameters were determined using the $\alpha$-enhanced \citet{Castelli2003} model atmospheres and the radiative transfer code \code{MOOG} including scattering opacity \citep{Sneden1973, Sobeck2011}\footnote{\url{https://github.com/alexji/moog17scat}} that assumes Local Thermodynamic Equilibrium (LTE).
Effective temperature ($\Teff$) was found by balancing Fe I abundance against excitation potential, surface gravity ($\logg$) was found by balancing Fe I and II abundances, and microturbulence ($\nu_t$) was found by balancing Fe II abundance vs reduced equivalent width.
We used Fe II abundances to determine microturbulence, both because they are safe from non-LTE effects, and because our good Fe II lines spanned a larger range of reduced equivalent widths compared to the Fe I lines in the observed red giants \citep{Ji2020b}.
Stellar parameter uncertainties were found by varying the stellar parameters to match the $1\sigma$ uncertainty on the standard error of the relevant slopes or abundance differences, adding a systematic uncertainty of 100\,K for $T_{\rm eff}$, 0.2 dex for $\log g$, 0.2 \kms for $\nu_t$, and 0.2 dex for [M/H] in quadrature.
Stellar parameters for the stars are found in Table~\ref{tab:obs}.

We derived stellar parameters assuming a spectroscopic effective temperature scale, but in general photometric temperatures tend to be a bit hotter \citep[e.g.][]{Frebel2013}. To estimate the difference, we applied the temperature correction in \citet{Frebel2013} and redetermined the stellar parameters following that paper, except for using Fe II abundances to determine microturbulence.
The \citet{Frebel2013} recalibrated temperatures were similar to the temperatures determined by a \emph{Gaia} color-temperature relation from a \code{PARSEC} isochrone \citep{Bressan2012} and \code{StarHorse} colors \citep{Anders2022}, except when the reddening correction was large.
Overall we found that the [Fe/H] of our Typhon stars increased by 0.1 dex when using photometric temperatures, but the [X/Fe] ratios stayed about the same. We thus decided to keep the spectroscopic stellar parameters, both to avoid uncertainties due to reddening and because it results in more accurate synthetic spectra models.

With these stellar parameters, we used synthetic spectra fits to determine the chemical abundances of ${\approx}50$ features that were heavily blended, required hyperfine or isotope splitting, and/or were molecular features. The best-fit abundances were found with $\chi^2$ minimization, marginalizing over a local continuum, resolution, and radial velocity shift as free parameters in addition to the element abundance (see \citealt{Ji2020b} for details).
For C-H and C-N, we adopted a ratio of 90\% $^{12}$C and 10\% $^{13}$C.
For Ba and Eu, we adopted solar $r$-process isotopic ratios \citep{Sneden2008}.
In total, 31 species of 26 elements were measured, where C-H, C-N, Al I, Sc II, V I and II, Mn I, Co I, Sr II, Y II, Zr II, Ba II, La II, Eu II, and Dy II were synthesized, as was the Si 3905{\AA} line; and abundances of all other features were determined with equivalent widths.
Individual line measurements and associated uncertainties are provided as supplemental material online.

Abundance averages and uncertainties were derived following \citet{Ji2020b}, which propagates the stellar parameter uncertainty as a weight to each individual line and properly accounts for correlations between line abundances due to stellar parameters.
We set the correlations between stellar parameters to zero in this analysis, which is closer to the uncertainty analysis in most literature sources we compare against. There were no significant differences in our conclusions adopting the stellar parameter correlations from \citet{Ji2020b}, though the detailed uncertainties and correlations are affected.
The final abundances are given in Table~\ref{tab:abunds}.

We briefly comment on a few specific stars and elements.
Star 3793377208170393984 had only Na D lines to measure sodium, and these were clearly blended with ISM features so we rejected them.
Since Na is strongly affected by non-local thermodynamic equilibrium (NLTE) effects, we have calculated NLTE corrections from \citet{Lind2011} and included both the LTE and NLTE [Na/Fe] abundances in Table~\ref{tab:abunds}. The uncertainty analysis is the same, so only the mean abundances change. We use the NLTE Na abundances in the rest of the text and figures.

We could detect the 3944{\AA} and 3961{\AA} Al lines in all stars, but due to heavy blends with Ca HK and other features we only included their abundance if the two lines gave a consistent result. The 6696{\AA} and 6698{\AA} Al lines were nonexistent in our spectra. The Al abundances should thus be considered with caution.
The C abundances are determined from the C-H bands around 4300{\AA}, assuming [O/Fe] = $+0.4$.
The O abundances are determined from the forbidden 6300{\AA} line, which is often blended with telluric lines and should be considered with caution. We have not provided O abundances if we suspected telluric contamination.
The Co abundances do not agree between different lines and should be considered unreliable unless investigating the individual line abundances provided.
Finally, we examined [X/Fe] as a function of $\Teff$, finding significant trends for C-H, Na I, Al I, Si I, and V I. The trend in C-H is expected as red giant branch stars convert C to N as they ascend the giant branch, but the others may suggest systematic effects (e.g. non-LTE effects) are affecting the results for those elements.

Star 3793377208170393984 has a higher $\Teff$ and $\nu_t$ than expected for its $\logg$ if it were to be a red giant branch star. This suggests it could be a red horizontal branch (RHB) star, which is corroborated by the rotational velocity from \code{Payne4MIKE} (5 \kms instead of unresolved $<2$ \kms for the other stars) and visually by its slightly wider lines (see \citealt{Roederer2014}). The color-magnitude diagram in \citet{Tenachi2022} also suggests this is a RHB star.
The LAMOST metallicities for this star \citep{Tenachi2022,Wang2022} have a higher $\mbox{[Fe/H]}=-1.25 \pm 0.09$, which could be due in part to its somewhat unusual evolutionary stage.

Star 3736372993468775424 has enhanced [Ba/Fe] compared to the other stars. Closer investigation suggests this star has likely experienced mass transfer from an AGB companion, as it also has relatively high [C/Fe] compared to other Typhon stars at its temperature.
We found no indication of velocity variations between \Gaia DR3 and our MIKE velocity.

\section{Results}\label{sec:results}

Table~\ref{tab:abunds} at the end of the paper gives the chemical abundances of the seven Typhon stars.
The $\log\epsilon$ abundances are the weighted average of different lines, where the weights account for the correlated effect of stellar parameters on all lines used to derive the abundance, by propagating the per-line abundance gradient with respect to each stellar parameter into a covariance matrix and taking the optimal estimator.
The [X/H] abundances are relative to \citet{Asplund2009} solar abundances.
$\sigma_{\rm [X/H]}$ is the total standard error on [X/H] including consistently propagating spectrum noise uncertainties, stellar parameter uncertainties from Table~\ref{tab:obs}, and an empirical systematic uncertainty to each line based on the line-to-line scatter.
[X/Fe] is calculated relative to [Fe I/H] if X is a neutral species and [Fe II/H] if X is an ionized species, while $\sigma_{\rm [X/Fe]}$ is the standard error also propagating stellar parameter uncertainties to both X and Fe I or II, which explains why its uncertainty differs from [X/H].
For more details see appendix B of \citet{Ji2020b}.

\subsection{Typhon is likely a tidally disrupted dwarf galaxy}
\begin{figure*}
    \centering
    \includegraphics[width=0.8\linewidth]{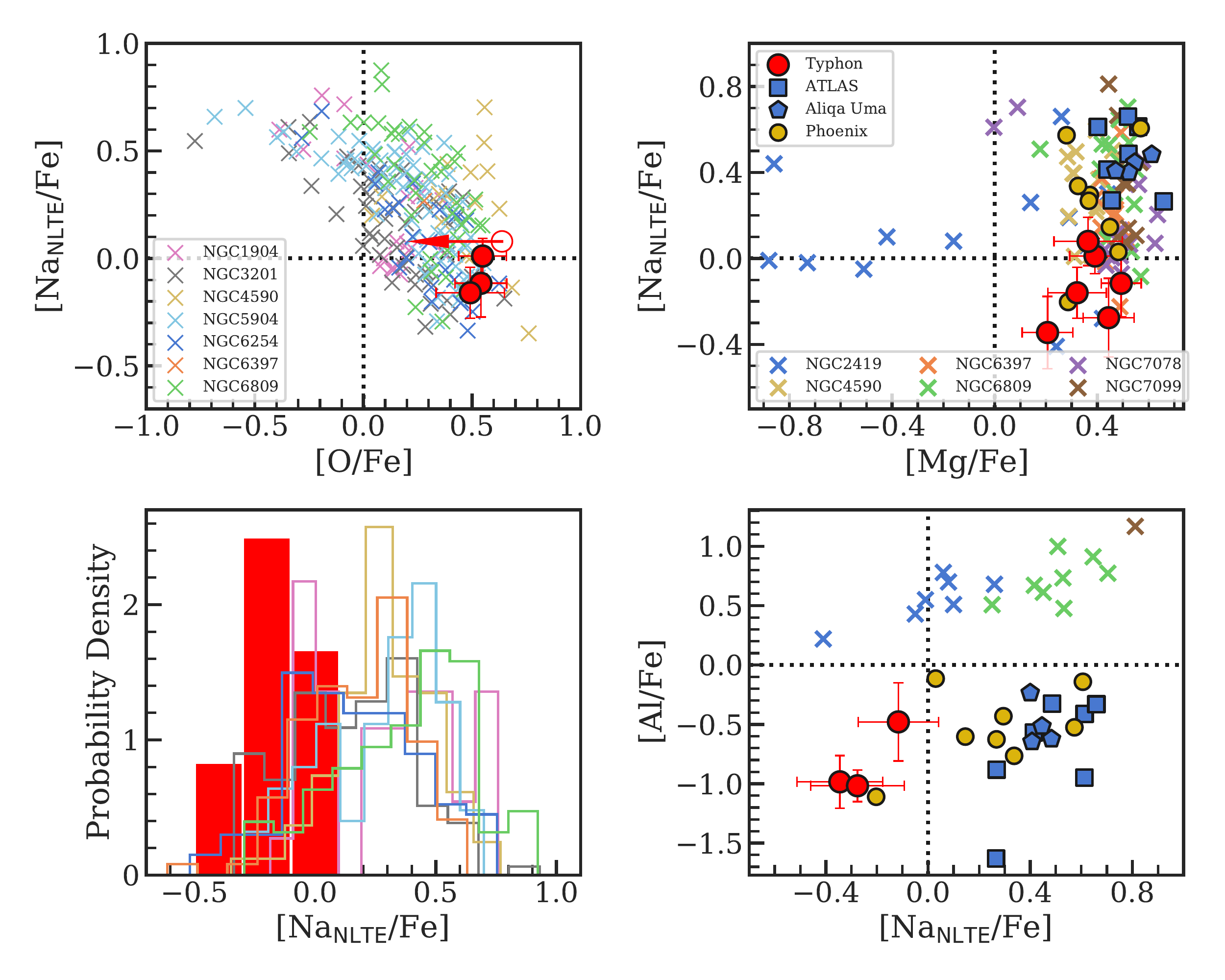}
    \caption{
    Typhon abundances (red) compared to globular cluster stars in the literature and S5 globular cluster streams \citep{Ji2020b}.
    The top-left and bottom-left panels show GC data with $-2.3 < \mbox{[Fe/H]} < -1.3$ from \citet{Carretta2009a} compared to Typhon.
    There is no characteristic Na-O anticorrelation observed in Typhon, though there are only three stars with both Na and O measurements.
    More important is the [Na/Fe] distribution, which displays no observable spread Typhon ($\sigma_{\rm Na} < 0.19$), in contrast to other globular clusters ($0.2 < \sigma_{\rm Na} < 0.3$).
    The top-right and bottom-right panels show two other element pairs that often show intrinsic dispersions in globular clusters \citep{Carretta2009b}.
    Though the number of abundance measurements is very limited, Typhon does not display any of the expected intrinsic spreads.
    }
    \label{fig:gccomp}
\end{figure*}

We first consider the metallicity dispersion of Typhon. Fitting a Gaussian metallicity distribution to all 7 stars with a log-uniform prior for $\sigma_{\rm Fe} \in [10^{-3}, 10^0]$, we infer $\langle\mbox{[Fe/H]}\rangle=-1.69 \pm 0.11$ with a metallicity dispersion of $\sigma_{\rm Fe}=0.28^{+0.11}_{-0.07}$.
Including the three LAMOST [Fe/H] measurements in \citet{Tenachi2022} gives essentially identical results, $\langle\mbox{[Fe/H]}\rangle=-1.67 \pm 0.09$ and $\sigma_{\rm Fe}=0.26^{+0.08}_{-0.06}$.
As found in \citet{Tenachi2022}, this dispersion depends entirely on the most metal-poor star in Typhon: removing that star results in an unresolved metallicity dispersion $\sigma_{\rm Fe} < 0.12$ at 95\% confidence, i.e. the six metal-rich stars in Typhon have no apparent [Fe/H] dispersion.
However, we agree with \citet{Tenachi2022} that there is no obvious reason to reject the metal-poor star as being a Typhon member, even though it is a spectroscopic binary.
While it may appear odd to have several stars in a metal-rich peak with one metal-poor star, we note that this is quite consistent with a leaky box metallicity distribution function, which is skewed-left in this way \citep{Kirby2013}.
More quantitatively, we considered the metallicity distribution for Leo I from \citet{Kirby2013}, which has the closest mean metallicity to Typhon. Leo I is best-fit by an Extra Gas model ($\langle\mbox{[Fe/H]}\rangle = -1.45$, $p_{\rm eff}=0.043$, $M=7.9$), where only 4.1\% of stars have $\mbox{[Fe/H]} < -2$, easily consistent with 1 out of 7 Typhon stars being metal-poor.
Thus we conclude that Typhon likely has a metallicity distribution consistent with a dwarf galaxy.

If the metal-poor star is an interloper and Typhon is a globular cluster, we might expect that Typhon displays anomalous chemical abundances indicative of multiple stellar populations \citep{Bastian2018,Gratton2019}.
Figure~\ref{fig:gccomp} shows the Na-O, Mg-Na, and Na-Al (anti-)correlations found in Galactic globular clusters \citep[e.g.][]{Yong2005,Carretta2009a,Carretta2009b}.
The Typhon stars do not display any light element scattering expected for multiple stellar populations.
This conclusion is based primarily on the lack of an observed Na dispersion in six Typhon stars (five excluding the metal-poor star), because only three stars have reasonable N or Al measurements and the abundance uncertainties on the even-$Z$ elements are relatively large compared to the size of the decrease.
Using the same Gaussian metallicity dispersion model as Fe, we find $\sigma$([Na$_{\rm NLTE}$/Fe]$<0.19$ at 95\% confidence, compared to typical dispersions of $0.2-0.3$ dex in other globular clusters \citep{Carretta2009b}. Furthermore, the top panels of Figure~\ref{fig:gccomp} show that the mean Na abundances are all similar to the primordial population in globular clusters.
Over 50\% of the stars in all of the Milky Way's nearby globular clusters have unusual chemical abundances that probably should be detected in these stars \citep{Milone2017}, so the naive probability of 0 out of 5 stars displaying a Na enhancement is $0.5^5 \lesssim 3\%$.
Typhon also shows no Mg-K dispersion as found in NGC2419 \citep{Cohen2012}.
However, the fraction of stars displaying unusual chemistry in accreted globular clusters may be substantially lower, possibly due to their lower initial stellar mass (\citealt{Milone2020}, Usman et al. in prep), so this is only a weak constraint.

Between the metallicity dispersion and lack of multiple populations, we conclude the evidence is in favor of Typhon not being a globular cluster, though future observations could disprove this.
Assuming Typhon is a new dwarf galaxy, using  $\langle\mbox{[Fe/H]}\rangle=-1.69 \pm 0.11$,
the mass-metallicity from \citet{Kirby2013} for intact dwarf galaxies gives
$\log M_\star/M_\odot = 6.00^{+0.54}_{-0.53}$.
The mass-metallicity relation from \citet{Naidu2022b} for tidally disrupted dwarf galaxies suggests Typhon should have had a stellar mass of $\log M_\star/M_\odot = 7.2^{+0.8}_{-0.8}$.
It is not clear which of these two relations should be used, since Typhon has a high apocenter and energy indicating it could have accreted onto the Milky Way relatively recently, but its high-$\alpha$ naively suggests it should have accreted relatively early.
Either way, the stellar mass is very high, comparable to some of the most massive intact dwarf galaxies around the Milky Way like Sculptor or Fornax, which suggests that there should be many more nearby Typhon members that have not yet been identified.

\subsection{Chemical evolution in Typhon}
\begin{figure*}
    \centering
    \includegraphics[width=\linewidth]{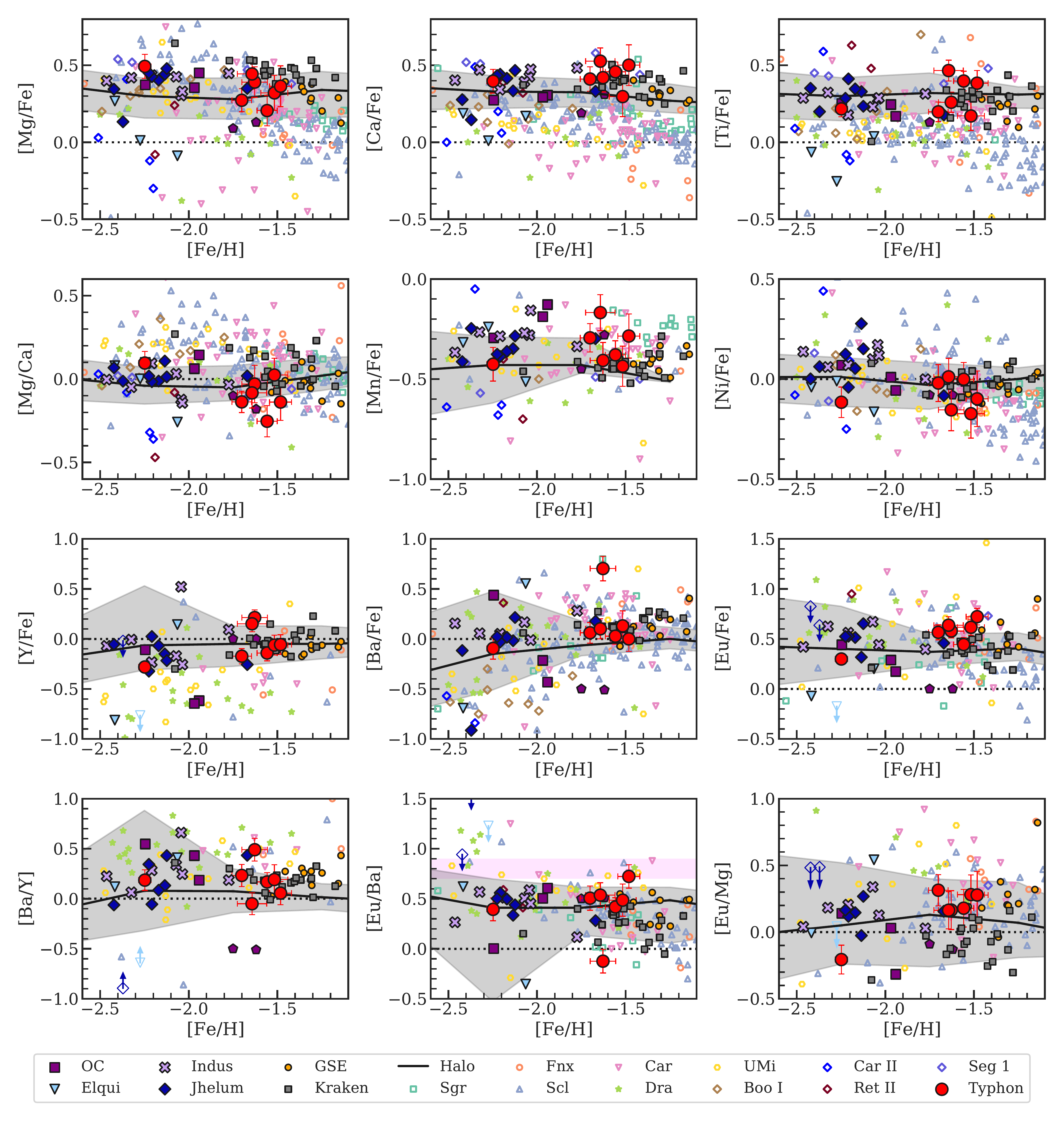}
    \caption{Chemical abundances of Typhon stars (large red circles) compared to stars in intact dwarf spheroidal galaxies, dwarf galaxy streams from $S^5$, and the massive halo structures GSE and Kraken (see text for references). Median [X/Y] abundances from stars in the Milky Way's stellar halo in bins of [Fe/H] are shown as a solid black line, with the shaded grey region indicating 68\% scatter. A horizontal dashed line is shown at [X/Y] $= 0$.
    The top row shows [$\alpha$/Fe], where Typhon has relatively high $\alpha$-element abundances.
    The middle row shows [Mg/Ca] and the Fe-peak elements, where Typhon mostly matches other dwarf galaxies.
    The third row shows several neutron-capture element abundances. The bottom row shows useful neutron-capture element ratios. The pink region for [Eu/Ba] indicates the pure $r$-process abundance ratio \citep{Sneden2008}. One star in Typhon is $s$-process enhanced, but the bulk of Typhon stars are somewhat $r$-process enhanced. The increased [Eu/Mg] at high [Fe/H] suggests Typhon has been enriched by delayed $r$-process sources, such as neutron star mergers.
    }
    \label{fig:dsphcomp}
\end{figure*}

Figure~\ref{fig:dsphcomp} shows chemical abundances of Typhon stars compared to different dwarf galaxies, both intact dSph galaxies \citep{Jonsson2018,Hansen2018,Letarte2010,Shetrone2003,Hill2019,Jablonka2015,Kirby2012,Norris2017,Cohen2009,Cohen2010,Tsujimoto2017} and dwarf galaxy streams observed by the Southern Stellar Stream Spectroscopic Survey \citep[$S^5$,][]{Ji2020b,Hansen2021,Li2022}, as well as abundances derived from MIKE observations of the Orphan Stream \citep{Casey2014}, and the disrupted dwarf galaxies Gaia-Sausage-Enceladus (GSE) and Kraken/Heracles \citep{Naidu2022}. We also plot halo stars from selected references in JINAbase for comparison as a shaded grey region \citep{Abohalima2018,Fulbright2000,Barklem2005,Cohen2013,Roederer2014,Jacobson2015}.
The top row shows [$\alpha$/Fe] abundances for Mg, Ca, and Ti; the second row shows [Mg/Ca] and Fe-peak abundances; and the bottom two rows show various combinations of neutron-capture elements Y, Ba, and Eu that trace the first, second, and rare-earth neutron-capture element peaks.
The Fe-peak abundances are overall similar to other dwarf galaxies and the stellar halo, with possibly the exception of [Mn/Fe] which is slightly enhanced and could be a useful chemical tag \citep{delosReyes2020}. We did not consider possible non-LTE effects for Mn \citep{Bergemann2019} so we refrain from interpreting it further.

The top row of Figure~\ref{fig:dsphcomp} shows the most important result of this study, which is that Typhon has elevated [$\alpha$/Fe] ratios in Mg and Ca.
We have tested a large number of permutations of stellar parameters and weighted averages to determine the mean abundance for each star, and this enhancement is very robust.
In particular, we have used the same line list and overall analysis method to derive the chemical abundances of stars in $S^5$ streams \citep{Ji2020b}, and it is clear that the most metal-rich Jhelum star (Jhelum\_2\_2) has similar [Mg/Fe] as but lower [Ca/Fe] than the typical Typhon star (Figure~\ref{fig:spectra}).
While there could be more metal-rich stars in Typhon that have not yet been found, at face value this suggests that Typhon quenched rather early in cosmic history, as it stopped forming stars while it was dominated by core-collapse instead of Type~Ia supernovae \citep[e.g.][]{Johnston2008,Lee2015,Cunningham2021}.
For comparison, the intact dSph with the closest [Fe/H] is Sculptor, with $\langle\mbox{[Fe/H]}\rangle=-1.7$ \citep{Kirby2013}, and Typhon has similar [Mg/Fe] and higher [Ca/Fe] than Sculptor \citep{Hill2019}.
Lower mass dwarf galaxies like Draco and Ursa Minor have similar or lower [$\alpha$/Fe] \citep{Norris2017,Cohen2009}.
Two other faint dwarf galaxy streams with lower apocenters, Jhelum and Indus, also display this high-$\alpha$ feature, with the same caveat that more metal-rich stars not yet studied in these streams may show lower-$\alpha$-element abundances \citep{Ji2020b,Hansen2021}.
In contrast, the recently accreted Orphan/Chenab (OC) stream displays quite low [$\alpha$/Fe] abundances at $\mbox{[Fe/H]}=-1.6$ \citep{Casey2014,Ji2020b,Hawkins2022,Naidu2022b}.

The Eu abundances in the bottom two rows of Figure~\ref{fig:dsphcomp} show an intriguing trend, namely that [Eu/Fe] appears to increase with [Fe/H] in Typhon.
The difference in [Eu/Fe] between the most Fe-poor star and the other Typhon stars is significant: like [Fe/H], the [Eu/Fe] dispersion is resolved when including the low-metallicity star, while it is unresolved when only considering the metal-rich Typhon stars.
This trend is opposite most dwarf galaxies, where [Eu/Fe] stays flat or declines at high metallicity, though a similar increasing [Eu/Fe] trend is also seen in Ursa Minor \citep{Cohen2010}.
An increasing [Eu/Fe] or [Eu/Mg] at higher [Fe/H] can indicate the onset of delayed $r$-process enrichment in a dwarf galaxy \citep[e.g.][]{Skuladottir2019,Aguado2021,Matsuno2021,Reggiani2021,Naidu2022}.
This is corroborated by the lower [Ba/Eu] in Typhon at high [Fe/H], indicating an increased $r$-process dominance at higher metallicity (except for the $s$-process mass transfer star, see Section~\ref{sec:methods}).
It is also possible such an enhancement could be due to extreme inhomogeneous metal mixing \citep[e.g.][]{Hansen2021,Reichert2021}, though Typhon's high mass inferred from the mass-metallicity relations ($10^6$ to $10^7 M_\odot$) suggests it is unlikely to have a lot of stochastic enrichment (\citealt{Naidu2022}, Frebel \& Ji submitted).
Note that there is a slight trend between \Teff and [Eu/Fe] for the high-Fe Typhon stars, so one should not over-interpret the slightly rising trend in [Eu/Fe] within the clump.

In summary, Typhon's chemical abundances somewhat conflictingly indicate a star formation history short enough to remain $\alpha$-enhanced, but long enough to be enriched by delayed $r$-process sources.

\section{Discussion and Conclusion}\label{sec:conclusion}

Our chemical evidence, along with the spatial and dynamics arguments by \citet{Tenachi2022}, suggests that Typhon is probably a dwarf galaxy.
The evidence primarily hinges on the membership of the most metal-poor star in Typhon and a small number of [Na/Fe] measurements, so we caution that future observations could end up showing this early conclusion to be incorrect.
However, assuming this star is a member, then Typhon is a dwarf galaxy with three somewhat contradictory observational facts:
\begin{enumerate}
    \item Typhon's metal-rich stars are moderately $\alpha$-enhanced, indicating it was forming stars efficiently at the end of its probably short life;
    \item Typhon's metal-rich stars show evidence for increased Eu-enhancement, indicating that it captured delayed $r$-process production from a source like neutron star mergers;
    \item Typhon has an apocenter larger than 100 kpc, usually interpreted as meaning it accreted relatively recently into the Milky Way. 
\end{enumerate}

\begin{figure*}
    \centering
    \includegraphics[width=\linewidth]{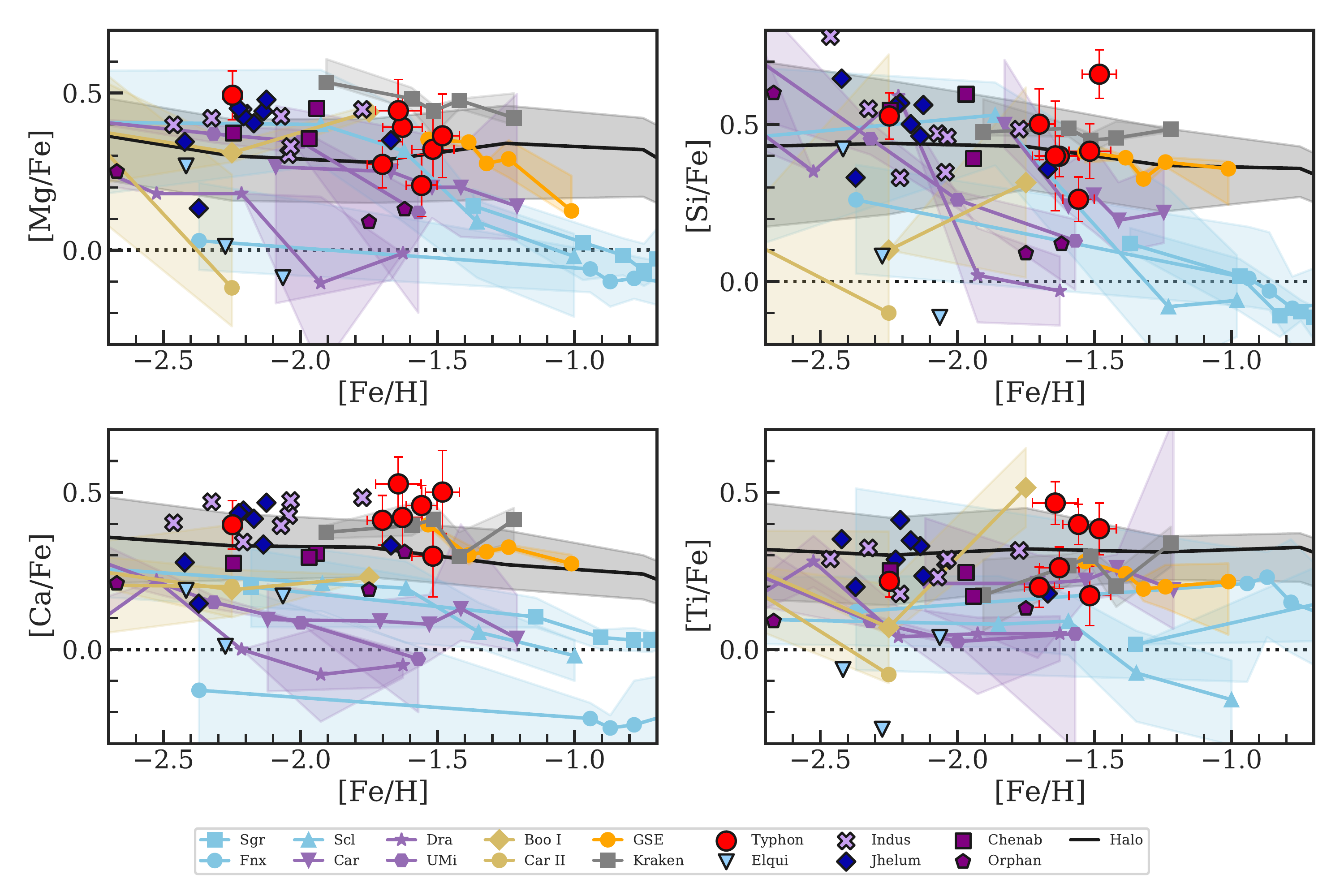}
    \caption{Detailed comparison of $\alpha$-element abundances between Typhon and other dwarf galaxies.
    To more clearly show the trends, data for the intact and fully disrupted dwarf galaxies have been placed into equal-star-number [Fe/H] bins, and only the median and $68\%$ scatter within each bin are plotted.
    The intact dSphs have been color coded according to their estimated stellar mass, with more massive dSphs in blue, intermediate mass dSphs in purple, and low mass dSphs in yellow.
    In general, Typhon is at the high end of [$\alpha$/Fe] compared to other dSphs. Galaxies overlapping with Typhon in [$\alpha$/Fe] continue forming stars to higher [Fe/H], while currently known Typhon members do not continue to higher metallicities.
    This would usually be interpreted as an abrupt truncation of Typhon's star formation, possibly because it accreted onto another galaxy and was quenched by environmental effects.
    }
    \label{fig:dsphalpha}
\end{figure*}

To elaborate, the typical explanation for high $\alpha$-enhancement in a satellite dwarf galaxy's most metal-rich stars is that it only formed stars for a short duration due to environmental quenching from early accretion onto the Milky Way galaxy \citep[e.g.][]{Johnston2008}. We show this in more detail in Figure~\ref{fig:dsphalpha}, where low-mass galaxies (purple) whose most metal-rich stars match Typhon's metal-rich stars have lower [$\alpha$/Fe]; and more massive galaxies (blue) have similar [$\alpha$/Fe] as Typhon but all continue to higher [Fe/H] while Typhon truncates.
If Typhon is a new dwarf galaxy, this explanation implies Typhon would have formed stars for a shorter duration than Sculptor, which has a similar [Fe/H] as Typhon but continues to evolve to higher [Fe/H] and lower [$\alpha$/Fe]. Sculptor formed 90\% of its stars within only the first 2\,Gyr of the universe according to \citet{Weisz2014}\footnote{\citet{delosReyes2022} argue that chemical evolution shows the total duration of star formation is ${\lesssim}1$\,Gyr, which could be consistent with \citet{Weisz2014} if the beginning of star formation was delayed; \citet{Hill2019} alternatively argue Sculptor formed its stars over 7\,Gyr using RGB ages.}.
The high [Mg/Fe] also matches other tidally disrupted halo dwarf galaxies that quenched at $z \approx 2-3$, when the universe is about 2-3 Gyr old \citep{Naidu2022b}.
However, such a short star formation duration is in tension with Typhon's high apocenter and $r$-process evolution.
For example, compare Typhon to Gaia--Sausage-Enceladus (GSE), which has the bulk of its stars with apocenters within 30-50kpc \citep{Deason2018,Naidu2021} and is certainly more massive than Typhon, but has a similar [Mg/Fe] as Typhon at $\mbox{[Fe/H]}=-1.6$ \citep[see Figure~\ref{fig:dsphalpha}]{Helmi2018,Haywood2018,Mackereth2019}.
Thus, at face value Typhon was forming stars as efficiently as GSE at the end of its life, but it has a larger apocenter that naively suggests it accreted onto the Milky Way (and thus quenched) at a later time, as well as an increasing [Eu/Fe] (and [Eu/Mg]) trend that indicates it captured delayed $r$-process element production \citep{Aguado2021,Matsuno2021,Naidu2022}.

We suggest four possible explanations to explain these apparently conflicting properties.

\emph{(1) Late-time dynamical interaction.} \citet{Tenachi2022} noted that Typhon may have had a past interaction with Sgr. Typhon could thus potentially have accreted early onto the Milky Way, shutting off its star formation before enough SN Ia enrichment occurred to affect [$\alpha$/Fe], and then was scattered to higher energies by a later interaction. This would require a stochastic explanation for the late-time $r$-process enhancement, and it can be tested by additional dynamical simulations. 

\emph{(2) Late-time star formation enhancement.} The metal-rich Typhon stars could have formed in a starburst that enhances [$\alpha$/Fe]. It is well-known that this ratio does not have to monotonically decline as a galaxy grows, but it can increase e.g. during a starburst that causes core-collapse supernovae to increase relative to Type~Ia supernovae \citep[e.g.][]{Weinberg2017}.
This has been seen in the Large Magellanic Cloud \citep{Nidever2020} and likely in the early Milky Way as well \citep{Conroy2022}.
With only one metal-poor star anchoring the tail of the chemical evolution history, and a large [Fe/H] gap between that star and the Fe-rich stars, Typhon's chemical evolution history could potentially display such a varying [$\alpha$/Fe] ratio, which can be tested with future chemical abundances.
Delayed star formation could also potentially explain the [Eu/Fe] rise at late times, since neutron star mergers produce Eu through the $r$-process and explode with a time delay. However, increasing Eu relative to Fe requires Eu to either increase faster than Fe, e.g. through a difference in the delay time distribution of neutron star mergers and Type~Ia supernovae.

\emph{(3) Early accretion and late-time group infall.} Typhon could have been first quenched by environmental interactions with a dwarf galaxy, and only later fell into the Milky Way. \citet{Wetzel2015} found in dark matter simulations that about $50\% \pm 10\%$ of \emph{intact} dwarf galaxies with stellar masses $10^6-10^7 M_\odot$ ($M_{\rm peak} \sim 10^{10} M_\odot$) have passed within the virial radius of another galaxy before falling into the Milky Way, though only 10-30\% of such systems remained as satellites at the time of infall.
Such an environmental interaction could have kept Typhon $\alpha$-enhanced by early environmental quenching, but allow it to have a large apocenter around the Milky Way due to a later infall.
The statistics of this scenario in cosmological simulations should be revisited in the context of tidally disrupted dwarf galaxies as well as in hydrodynamic simulations.

\emph{(4) High-energy tail of another disrupted galaxy.} Dwarf galaxies accreting into the Milky Way do not deposit all their material at a single energy, as dynamical friction causes them to sink to lower energies (and thus smaller apocenters) with each orbit \citep[e.g.][]{Naidu2021}. 
Thus, while no known structure resides in the same kinematic space as Typhon \citep{Tenachi2022,Dodd2022},
Typhon could still potentially be the first orbital wrap of another accretion event. Such an event would likely strip only a relatively metal-poor part of the galaxy, such that our discussion does not reflect the whole galaxy's chemical evolution \citep[e.g.][]{Johnson2020, Naidu2021}.

This scenario would likely require Typhon being the tail of a massive accretion event, as massive galaxies are most strongly affected by dynamical friction. Thus, we probably should already know the possible progenitiors of Typhon in this scenario.
Examining all known large accretion events \citep{Naidu2020,Bonaca2021,Li2022,Malhan2022},
we reject the three most massive accretion events as possible origins of Typhon:
the LMC pericenter of ${\sim}50\,$kpc is not nearly small enough to match Typhon's 6\,kpc pericenter \citep{Tenachi2022};
GSE's high-energy debris should be much more retrograde than Typhon \citep{Naidu2021};
and Sgr has the wrong sign for $L_y$.

There are also myriad known closer dwarf galaxy streams with moderately high metallicities that are similar or slightly more massive than Typhon (e.g. Palca/Cetus, LMS-1/Wukong, the Helmi stream, and others; \citealt{Naidu2020,Bonaca2021,Li2022,Malhan2022}). The initial infall of these systems has not yet been studied in fully live N-body simulations, so it remains possible that Typhon is an early wrap of one of these systems.
A path forward to rule out such associations is to compare the detailed chemical abundances of these nearby lower-mass streams to Typhon. Palca/Cetus does not have any detailed chemical abundances published. LMS-1/Wukong has a similar [Mg/Fe] as Typhon in APOGEE and H3, but more elements are needed to compare with Typhon \citep{Horta2022,Naidu2020}. The Helmi stream has been chemically studied many times \citep{Roederer2010,Aguado2021b,Gull2021,Limberg2021,Matsuno2022}.
\citet{Limberg2021} summarized these results, including a crucial re-assessment of stream membership with \Gaia, which removed a large number of stars previously hypothesized to be part of the Helmi stream. At $\mbox{[Fe/H]}=-1.7$, they find the Helmi stream has similar [Mg/Fe] and [Eu/Fe] but lower [Ca/Fe] than Typhon, so it is probably not the progenitor of Typhon though more data is certainly needed to be sure. 
\citet{Matsuno2022} also found a lower [Ca/Fe] in the Helmi stream, as well as a somewhat low [Y/Fe] ratio also inconsistent with Typhon.
This exercise emphasizes the crucial importance of pinning down accurate chemical signatures of metal-poor stars in \emph{known} accretion events before trying to discover new faint stellar streams.

\vspace{3mm}

To set an initial prior on whether Typhon is a bona fide dwarf galaxy (scenarios 1-3) or an early wrap of a larger dwarf galaxy (scenario 4), we examine the accreted halo star catalog from the \emph{Caterpillar} dark-matter-only simulations \citep{Griffen2016,Brauer2022}.
Briefly, halo star particles are identified by examining the merger tree of each Milky Way mass host halo to find all branches where a halo merged into the Milky Way. The 10\% most bound particles of each merging subhalo are tagged as stars at the time of peak mass and then traced to $z=0$. The orbital parameters and apocenters are computed using the \code{AGAMA} software library \citep{Vasiliev19} following \citet{Brauer2022}.
We then select halo star particles within a 5 kpc radius of the solar neighborhood at $z=0$, pick the subset with apocenters $> 100$\,kpc, and connect those star particles back to their original birth galaxies.
Fewer than 0.1\% of the tagged star particles have apocenters $> 100$\,kpc.
Using the \citet{Garrison-Kimmel2017} abundance matching relation, we find that 56\% of these high-apocenter solar neighborhood stars originated in massive galaxies with $M_\star > 10^{7.2} M_\odot$, 24\% of the stars originate in galaxies between $10^{6-7.2} M_\odot$, and 20\% of the stars originate in galaxies less than $10^6 M_\odot$ (which is disfavored for Typhon).
Note that the star particle method tags a constant fraction of dark matter particles, so the number of actual stars in lower mass galaxies is even lower due to the steepness of the stellar mass-halo mass relation.
In conclusion, ${\gtrsim}70\%$ of the time, high-apocenter stars with Typhon's metallicity are the tails of massive accretion events, which gives $70\%/30\% \approx 2:1$ odds that Typhon is \emph{not} an independent dwarf galaxy.
More detailed future investigation of this question would be beneficial, such as accounting for the shared kinematics and orbital phase of Typhon stars.

\vspace{0.3cm}
We also note that \citet{Yu2020} proposed an \emph{in situ} formation scenario that could explain Typhon-like systems. They found that very bursty feedback in the FIRE-2 simulations sometimes deposited a large amount of momentum into giant molecular clouds, triggering star formation and producing stars with halo-like orbits. The chemical abundances of Typhon are only moderately different from in situ metal-poor Milky Way stars, so we cannot rule out such a scenario.

\vspace{0.3cm}
Typhon emphasizes the need for a global picture to tackle the chemical and dynamical complexity expected for the Milky Way's tidally disrupted dwarf galaxy satellites.
If Typhon is the high-energy wrap of a massive dwarf galaxy, it shows the crucial importance of connecting apparently disparate kinematic components with detailed chemical abundances and dynamical models.
If instead Typhon is indeed a new dwarf galaxy, its different and unusual formation scenarios can likely be quickly tested with additional chemical abundance measurements within Typhon, especially because its high inferred stellar mass suggests many other Typhon stars should be nearby to be discovered.
Either way, simulations suggest that $50-80\%$ of dwarf galaxies that existed in the Milky Way volume at high redshift are now tidally disrupted \citep{Safarzadeh2018,Santistevan2020}, so the coming years are an exciting time for \Gaia and spectroscopic surveys to continue to unravel our Milky Way stellar halo into its component parts.

\begin{deluxetable}{lccrrrrr}
\centering
\tablewidth{\linewidth}
\tablecolumns{8}
\tabletypesize{\scriptsize}
\tablecaption{\label{tab:abunds}Stellar Abundances}
\tablehead{El. & $N$ & ul & $\log \epsilon$ & [X/H] & $\sigma_{\text{[X/H]}}$ & [X/Fe] & $\sigma_{\text{[X/Fe]}}$}
\startdata
\hline\multicolumn{8}{c}{Gaia DR3 1765600930139450752 }\\\hline
C-H   &   2 &     &$+6.12$&$-2.31$&  0.15 &$-0.07$&  0.14 \\
C-N   &   1 & $<$ &$+6.08$&$-1.75$&\nodata&$+0.49$&\nodata\\
O I   &   1 &     &$+6.98$&$-1.71$&  0.13 &$+0.54$&  0.12 \\
Na I  &   4 &     &$+4.01$&$-2.23$&  0.16 &$+0.02$&  0.16 \\
Na NLTE & & & $+3.88$&$-2.36$&&$-0.12$&\\
Mg I  &   2 &     &$+5.84$&$-1.75$&  0.08 &$+0.49$&  0.08 \\
Al I  &   2 &     &$+3.72$&$-2.73$&  0.33 &$-0.48$&  0.33 \\
Si I  &   6 &     &$+5.79$&$-1.72$&  0.07 &$+0.53$&  0.07 \\
K I   &   2 &     &$+3.54$&$-1.49$&  0.11 &$+0.76$&  0.11 \\
Ca I  &  12 &     &$+4.49$&$-1.85$&  0.08 &$+0.40$&  0.08 \\
Sc II &   7 &     &$+0.85$&$-2.30$&  0.08 &$-0.04$&  0.05 \\
Ti I  &  28 &     &$+2.92$&$-2.03$&  0.10 &$+0.22$&  0.10 \\
Ti II &  24 &     &$+2.91$&$-2.04$&  0.09 &$+0.22$&  0.05 \\
V I   &   2 &     &$+1.43$&$-2.50$&  0.06 &$-0.25$&  0.07 \\
V II  &   2 &     &$+1.73$&$-2.20$&  0.10 &$+0.06$&  0.06 \\
Cr I  &  13 &     &$+3.23$&$-2.41$&  0.13 &$-0.16$&  0.12 \\
Cr II &   5 &     &$+3.47$&$-2.17$&  0.09 &$+0.09$&  0.05 \\
Mn I  &   5 &     &$+2.75$&$-2.67$&  0.08 &$-0.43$&  0.08 \\
Fe I  &  78 &     &$+5.25$&$-2.25$&  0.04 &$+0.00$&  0.00 \\
Fe II &  19 &     &$+5.24$&$-2.26$&  0.08 &$+0.00$&  0.00 \\
Co I  &   5 &     &$+2.76$&$-2.23$&  0.12 &$+0.02$&  0.12 \\
Ni I  &  20 &     &$+3.86$&$-2.36$&  0.08 &$-0.12$&  0.08 \\
Cu I  &   1 &     &$+1.16$&$-3.03$&  0.16 &$-0.79$&  0.15 \\
Zn I  &   2 &     &$+2.41$&$-2.15$&  0.07 &$+0.10$&  0.07 \\
Sr II &   2 &     &$+0.76$&$-2.11$&  0.10 &$+0.15$&  0.10 \\
Y II  &   7 &     &$-0.33$&$-2.54$&  0.09 &$-0.28$&  0.05 \\
Zr II &   1 &     &$+0.36$&$-2.22$&  0.10 &$+0.04$&  0.05 \\
Ba II &   4 &     &$-0.17$&$-2.35$&  0.13 &$-0.10$&  0.11 \\
La II &   5 &     &$-1.10$&$-2.20$&  0.08 &$+0.06$&  0.04 \\
Eu II &   3 &     &$-1.44$&$-1.96$&  0.09 &$+0.30$&  0.06 \\
Dy II &   2 &     &$-0.79$&$-1.89$&  0.09 &$+0.37$&  0.04 \\
\hline\multicolumn{8}{c}{Gaia DR3  3573787693673899520 }\\\hline
C-H   &   2 &     &$+6.73$&$-1.70$&  0.11 &$-0.14$&  0.11 \\
C-N   &   1 & $<$ &$+7.51$&$-0.32$&\nodata&$+1.24$&\nodata\\
O I   &   1 & $<$ &$+8.86$&$+0.17$&\nodata&$+1.73$&\nodata\\
Na I  &   2 &     &$+4.87$&$-1.37$&  0.18 &$+0.19$&  0.17 \\
Na NLTE & & & $+4.34$&$-1.90$&&$-0.35$&\\
Mg I  &   2 &     &$+6.25$&$-1.35$&  0.10 &$+0.21$&  0.10 \\
Al I  &   2 &     &$+3.91$&$-2.54$&  0.23 &$-0.98$&  0.22 \\
Si I  &   4 &     &$+6.21$&$-1.30$&  0.06 &$+0.26$&  0.07 \\
K I   &   2 &     &$+4.30$&$-0.73$&  0.14 &$+0.82$&  0.13 \\
Ca I  &  12 &     &$+5.24$&$-1.10$&  0.06 &$+0.46$&  0.06 \\
Sc II &   8 &     &$+1.67$&$-1.48$&  0.08 &$+0.13$&  0.06 \\
Ti I  &  20 &     &$+3.61$&$-1.34$&  0.11 &$+0.22$&  0.10 \\
Ti II &  16 &     &$+3.73$&$-1.22$&  0.10 &$+0.40$&  0.06 \\
V I   &   2 &     &$+2.34$&$-1.59$&  0.12 &$-0.04$&  0.10 \\
V II  &   2 &     &$+2.49$&$-1.44$&  0.09 &$+0.18$&  0.07 \\
Cr I  &  10 &     &$+3.98$&$-1.66$&  0.11 &$-0.10$&  0.09 \\
Cr II &   5 &     &$+4.18$&$-1.46$&  0.09 &$+0.16$&  0.05 \\
Mn I  &   5 &     &$+3.49$&$-1.94$&  0.07 &$-0.38$&  0.07 \\
Fe I  &  80 &     &$+5.94$&$-1.56$&  0.06 &$+0.00$&  0.00 \\
Fe II &  16 &     &$+5.88$&$-1.62$&  0.09 &$+0.00$&  0.00 \\
Co I  &   3 &     &$+3.34$&$-1.65$&  0.11 &$-0.09$&  0.10 \\
Ni I  &   8 &     &$+4.66$&$-1.56$&  0.09 &$-0.00$&  0.08 \\
Zn I  &   2 &     &$+3.20$&$-1.36$&  0.08 &$+0.20$&  0.08 \\
Sr II &   2 &     &$+1.52$&$-1.35$&  0.12 &$+0.26$&  0.13 \\
Y II  &   3 &     &$+0.45$&$-1.76$&  0.10 &$-0.14$&  0.08 \\
Zr II &   1 &     &$+1.29$&$-1.29$&  0.09 &$+0.33$&  0.07 \\
Ba II &   4 &     &$+0.59$&$-1.59$&  0.10 &$+0.03$&  0.08 \\
La II &   3 &     &$-0.26$&$-1.36$&  0.10 &$+0.26$&  0.09 \\
Eu II &   2 &     &$-0.65$&$-1.17$&  0.09 &$+0.44$&  0.07 \\
Dy II &   1 &     &$-0.18$&$-1.28$&  0.18 &$+0.34$&  0.18 \\
\hline\multicolumn{8}{c}{Gaia DR3  3736372993468775424 }\\\hline
C-H   &   2 &     &$+6.99$&$-1.44$&  0.15 &$+0.18$&  0.13 \\
C-N   &   1 &     &$+6.09$&$-1.74$&  0.21 &$-0.11$&  0.18 \\
O I   &   1 &     &$+7.61$&$-1.08$&  0.11 &$+0.55$&  0.11 \\
Na I  &   4 &     &$+4.66$&$-1.58$&  0.07 &$+0.05$&  0.08 \\
Na NLTE & & & $+4.62$&$-1.62$&&$+0.01$&\\
Mg I  &   4 &     &$+6.36$&$-1.24$&  0.10 &$+0.39$&  0.10 \\
Si I  &   5 &     &$+6.28$&$-1.23$&  0.06 &$+0.40$&  0.07 \\
K I   &   2 &     &$+4.44$&$-0.59$&  0.20 &$+1.04$&  0.19 \\
Ca I  &  12 &     &$+5.13$&$-1.21$&  0.10 &$+0.42$&  0.09 \\
Sc II &   7 &     &$+1.57$&$-1.58$&  0.10 &$+0.07$&  0.07 \\
Ti I  &  36 &     &$+3.61$&$-1.34$&  0.12 &$+0.29$&  0.12 \\
Ti II &  18 &     &$+3.56$&$-1.39$&  0.09 &$+0.26$&  0.07 \\
V I   &   2 &     &$+2.06$&$-1.87$&  0.13 &$-0.24$&  0.15 \\
V II  &   1 &     &$+2.62$&$-1.31$&  0.11 &$+0.34$&  0.07 \\
Cr I  &  13 &     &$+3.85$&$-1.79$&  0.17 &$-0.16$&  0.15 \\
Cr II &   3 &     &$+4.09$&$-1.55$&  0.10 &$+0.10$&  0.06 \\
Mn I  &   5 &     &$+3.40$&$-2.04$&  0.09 &$-0.41$&  0.09 \\
Fe I  &  77 &     &$+5.87$&$-1.63$&  0.07 &$+0.00$&  0.00 \\
Fe II &  21 &     &$+5.85$&$-1.65$&  0.10 &$+0.00$&  0.00 \\
Co I  &   5 &     &$+3.37$&$-1.62$&  0.16 &$+0.01$&  0.16 \\
Ni I  &  22 &     &$+4.44$&$-1.78$&  0.11 &$-0.15$&  0.10 \\
Cu I  &   1 &     &$+2.28$&$-1.91$&  0.16 &$-0.29$&  0.14 \\
Zn I  &   2 &     &$+3.03$&$-1.53$&  0.10 &$+0.10$&  0.11 \\
Sr II &   2 &     &$+1.59$&$-1.28$&  0.05 &$+0.37$&  0.09 \\
Y II  &   6 &     &$+0.77$&$-1.44$&  0.10 &$+0.21$&  0.08 \\
Zr II &   1 &     &$+1.26$&$-1.32$&  0.18 &$+0.33$&  0.14 \\
Ba II &   4 &     &$+1.23$&$-0.95$&  0.09 &$+0.70$&  0.12 \\
La II &   5 &     &$+0.11$&$-0.99$&  0.10 &$+0.66$&  0.07 \\
Eu II &   3 &     &$-0.55$&$-1.07$&  0.08 &$+0.58$&  0.07 \\
Dy II &   3 &     &$+0.31$&$-0.79$&  0.09 &$+0.86$&  0.07 \\
\hline\multicolumn{8}{c}{Gaia DR3  3793377208170393984 }\\\hline
C-H   &   2 &     &$+6.41$&$-2.02$&  0.15 &$-0.32$&  0.14 \\
C-N   &   1 & $<$ &$+7.63$&$-0.20$&\nodata&$+1.50$&\nodata\\
O I   &   1 & $<$ &$+8.06$&$-0.63$&\nodata&$+1.07$&\nodata\\
Mg I  &   2 &     &$+6.17$&$-1.43$&  0.07 &$+0.27$&  0.07 \\
Al I  &   2 &     &$+3.72$&$-2.73$&  0.07 &$-1.03$&  0.09 \\
Si I  &   2 &     &$+6.31$&$-1.20$&  0.12 &$+0.50$&  0.11 \\
K I   &   2 &     &$+4.12$&$-0.91$&  0.12 &$+0.79$&  0.12 \\
Ca I  &  10 &     &$+5.05$&$-1.29$&  0.08 &$+0.41$&  0.08 \\
Sc II &   8 &     &$+1.61$&$-1.54$&  0.09 &$+0.14$&  0.08 \\
Ti I  &  10 &     &$+3.52$&$-1.43$&  0.12 &$+0.27$&  0.11 \\
Ti II &  14 &     &$+3.47$&$-1.48$&  0.11 &$+0.20$&  0.06 \\
V I   &   2 &     &$+2.17$&$-1.76$&  0.13 &$-0.06$&  0.12 \\
V II  &   2 &     &$+2.36$&$-1.57$&  0.11 &$+0.11$&  0.07 \\
Cr I  &   9 &     &$+3.75$&$-1.89$&  0.12 &$-0.19$&  0.11 \\
Cr II &   4 &     &$+3.91$&$-1.73$&  0.10 &$-0.05$&  0.06 \\
Mn I  &   5 &     &$+3.44$&$-1.99$&  0.07 &$-0.29$&  0.07 \\
Fe I  &  68 &     &$+5.80$&$-1.70$&  0.06 &$+0.00$&  0.00 \\
Fe II &  17 &     &$+5.82$&$-1.68$&  0.10 &$+0.00$&  0.00 \\
Co I  &   4 &     &$+3.37$&$-1.62$&  0.16 &$+0.08$&  0.15 \\
Ni I  &  12 &     &$+4.50$&$-1.72$&  0.10 &$-0.02$&  0.09 \\
Zn I  &   2 &     &$+2.91$&$-1.65$&  0.09 &$+0.05$&  0.08 \\
Sr II &   2 &     &$+1.34$&$-1.53$&  0.16 &$+0.15$&  0.14 \\
Y II  &   7 &     &$+0.36$&$-1.85$&  0.09 &$-0.17$&  0.08 \\
Zr II &   1 &     &$+0.96$&$-1.62$&  0.10 &$+0.06$&  0.06 \\
Ba II &   4 &     &$+0.56$&$-1.62$&  0.10 &$+0.06$&  0.10 \\
La II &   5 &     &$-0.39$&$-1.49$&  0.10 &$+0.19$&  0.07 \\
Eu II &   2 &     &$-0.59$&$-1.11$&  0.12 &$+0.57$&  0.07 \\
Dy II &   2 &     &$+0.17$&$-0.93$&  0.12 &$+0.75$&  0.09 \\
\hline\multicolumn{8}{c}{Gaia DR3  3891712266823336192 }\\\hline
C-H   &   2 &     &$+6.49$&$-1.94$&  0.17 &$-0.43$&  0.18 \\
C-N   &   1 &     &$+6.70$&$-1.13$&  0.28 &$+0.39$&  0.27 \\
O I   &   1 &     &$+7.67$&$-1.02$&  0.13 &$+0.49$&  0.16 \\
Na I  &   4 &     &$+4.62$&$-1.62$&  0.11 &$-0.11$&  0.12 \\
Na NLTE & & & $+4.56$&$-1.68$&&$-0.16$&\\
Mg I  &   2 &     &$+6.40$&$-1.20$&  0.12 &$+0.32$&  0.11 \\
Si I  &   4 &     &$+6.41$&$-1.10$&  0.07 &$+0.41$&  0.09 \\
K I   &   2 &     &$+4.50$&$-0.53$&  0.24 &$+0.99$&  0.23 \\
Ca I  &  10 &     &$+5.12$&$-1.22$&  0.13 &$+0.30$&  0.13 \\
Sc II &   4 &     &$+1.71$&$-1.44$&  0.14 &$+0.09$&  0.10 \\
Ti I  &  31 &     &$+3.60$&$-1.34$&  0.12 &$+0.17$&  0.12 \\
Ti II &  17 &     &$+3.59$&$-1.36$&  0.14 &$+0.17$&  0.10 \\
V I   &   2 &     &$+2.13$&$-1.80$&  0.16 &$-0.28$&  0.15 \\
V II  &   2 &     &$+2.42$&$-1.50$&  0.13 &$+0.03$&  0.10 \\
Cr I  &  12 &     &$+3.83$&$-1.80$&  0.21 &$-0.29$&  0.19 \\
Cr II &   5 &     &$+4.12$&$-1.52$&  0.13 &$+0.01$&  0.09 \\
Mn I  &   5 &     &$+3.48$&$-1.95$&  0.10 &$-0.44$&  0.10 \\
Fe I  &  76 &     &$+5.98$&$-1.52$&  0.08 &$+0.00$&  0.00 \\
Fe II &  20 &     &$+5.97$&$-1.53$&  0.12 &$+0.00$&  0.00 \\
Co I  &   5 &     &$+3.57$&$-1.42$&  0.15 &$+0.10$&  0.16 \\
Ni I  &  19 &     &$+4.53$&$-1.69$&  0.12 &$-0.17$&  0.12 \\
Cu I  &   1 &     &$+2.35$&$-1.84$&  0.20 &$-0.33$&  0.19 \\
Zn I  &   2 &     &$+3.08$&$-1.48$&  0.10 &$+0.04$&  0.13 \\
Sr II &   2 &     &$+1.53$&$-1.34$&  0.08 &$+0.19$&  0.10 \\
Y II  &   7 &     &$+0.62$&$-1.59$&  0.12 &$-0.06$&  0.10 \\
Zr II &   1 &     &$+1.27$&$-1.31$&  0.14 &$+0.23$&  0.11 \\
Ba II &   4 &     &$+0.78$&$-1.40$&  0.16 &$+0.13$&  0.15 \\
La II &   5 &     &$-0.06$&$-1.16$&  0.12 &$+0.37$&  0.08 \\
Eu II &   3 &     &$-0.40$&$-0.92$&  0.09 &$+0.62$&  0.08 \\
Dy II &   3 &     &$+0.21$&$-0.89$&  0.09 &$+0.65$&  0.11 \\
\hline\multicolumn{8}{c}{Gaia DR3  3913243629368310912 }\\\hline
C-H   &   2 &     &$+6.73$&$-1.70$&  0.17 &$-0.22$&  0.18 \\
C-N   &   1 &     &$+6.52$&$-1.31$&  0.25 &$+0.17$&  0.25 \\
O I   &   1 & $<$ &$+7.84$&$-0.84$&\nodata&$+0.64$&\nodata\\
Na I  &   4 &     &$+4.92$&$-1.32$&  0.11 &$+0.16$&  0.11 \\
Na NLTE & & & $+4.84$&$-1.40$&&$+0.08$&\\
Mg I  &   3 &     &$+6.48$&$-1.12$&  0.14 &$+0.36$&  0.13 \\
Si I  &   4 &     &$+6.69$&$-0.82$&  0.06 &$+0.66$&  0.08 \\
K I   &   2 &     &$+4.33$&$-0.70$&  0.21 &$+0.78$&  0.20 \\
Ca I  &  12 &     &$+5.36$&$-0.98$&  0.13 &$+0.50$&  0.13 \\
Sc II &   6 &     &$+1.90$&$-1.25$&  0.09 &$+0.32$&  0.11 \\
Ti I  &  26 &     &$+3.88$&$-1.07$&  0.14 &$+0.41$&  0.14 \\
Ti II &  18 &     &$+3.77$&$-1.18$&  0.12 &$+0.38$&  0.08 \\
V I   &   2 &     &$+2.31$&$-1.62$&  0.19 &$-0.14$&  0.18 \\
Cr I  &  14 &     &$+4.09$&$-1.55$&  0.21 &$-0.07$&  0.20 \\
Cr II &   2 &     &$+4.32$&$-1.32$&  0.18 &$+0.25$&  0.14 \\
Mn I  &   5 &     &$+3.66$&$-1.77$&  0.11 &$-0.28$&  0.11 \\
Fe I  &  61 &     &$+6.02$&$-1.48$&  0.06 &$+0.00$&  0.00 \\
Fe II &  20 &     &$+5.94$&$-1.56$&  0.13 &$+0.00$&  0.00 \\
Co I  &   5 &     &$+3.60$&$-1.39$&  0.19 &$+0.09$&  0.19 \\
Ni I  &  20 &     &$+4.64$&$-1.58$&  0.14 &$-0.10$&  0.14 \\
Cu I  &   1 &     &$+2.39$&$-1.80$&  0.19 &$-0.32$&  0.18 \\
Zn I  &   2 &     &$+3.20$&$-1.36$&  0.11 &$+0.12$&  0.13 \\
Sr II &   2 &     &$+1.61$&$-1.26$&  0.13 &$+0.30$&  0.15 \\
Y II  &   7 &     &$+0.59$&$-1.62$&  0.10 &$-0.06$&  0.10 \\
Zr II &   1 &     &$+1.16$&$-1.42$&  0.14 &$+0.14$&  0.10 \\
Ba II &   4 &     &$+0.62$&$-1.56$&  0.13 &$-0.00$&  0.12 \\
La II &   4 &     &$+0.05$&$-1.05$&  0.12 &$+0.51$&  0.10 \\
Eu II &   3 &     &$-0.32$&$-0.84$&  0.09 &$+0.72$&  0.08 \\
Dy II &   3 &     &$+0.28$&$-0.82$&  0.06 &$+0.74$&  0.11 \\
\hline\multicolumn{8}{c}{Gaia DR3  3939346894405032576 }\\\hline
C-H   &   2 &     &$+6.80$&$-1.63$&  0.17 &$+0.01$&  0.15 \\
C-N   &   1 & $<$ &$+7.56$&$-0.27$&\nodata&$+1.37$&\nodata\\
O I   &   1 & $<$ &$+8.62$&$-0.07$&\nodata&$+1.57$&\nodata\\
Na I  &   2 &     &$+4.89$&$-1.35$&  0.21 &$+0.29$&  0.18 \\
Na NLTE & & & $+4.32$&$-1.92$&&$-0.28$&\\
Mg I  &   2 &     &$+6.40$&$-1.20$&  0.10 &$+0.44$&  0.10 \\
Al I  &   2 &     &$+3.79$&$-2.66$&  0.15 &$-1.02$&  0.13 \\
Si I  &   1 &     &$+6.27$&$-1.24$&  0.21 &$+0.40$&  0.17 \\
K I   &   2 &     &$+4.33$&$-0.70$&  0.20 &$+0.95$&  0.19 \\
Ca I  &   8 &     &$+5.22$&$-1.12$&  0.09 &$+0.53$&  0.09 \\
Sc II &   6 &     &$+1.65$&$-1.50$&  0.10 &$+0.17$&  0.06 \\
Ti I  &   9 &     &$+3.81$&$-1.14$&  0.15 &$+0.50$&  0.13 \\
Ti II &  13 &     &$+3.74$&$-1.21$&  0.11 &$+0.47$&  0.07 \\
V I   &   1 &     &$+2.55$&$-1.38$&  0.11 &$+0.27$&  0.10 \\
V II  &   2 &     &$+2.55$&$-1.38$&  0.11 &$+0.29$&  0.10 \\
Cr I  &   4 &     &$+3.91$&$-1.73$&  0.16 &$-0.09$&  0.13 \\
Cr II &   3 &     &$+4.04$&$-1.60$&  0.09 &$+0.07$&  0.04 \\
Mn I  &   5 &     &$+3.62$&$-1.81$&  0.10 &$-0.17$&  0.09 \\
Fe I  &  51 &     &$+5.86$&$-1.64$&  0.08 &$+0.00$&  0.00 \\
Fe II &  11 &     &$+5.83$&$-1.67$&  0.09 &$+0.00$&  0.00 \\
Co I  &   4 &     &$+3.34$&$-1.65$&  0.12 &$-0.01$&  0.11 \\
Ni I  &   7 &     &$+4.59$&$-1.63$&  0.10 &$+0.01$&  0.09 \\
Zn I  &   2 &     &$+3.03$&$-1.53$&  0.11 &$+0.11$&  0.10 \\
Sr II &   2 &     &$+1.27$&$-1.60$&  0.21 &$+0.08$&  0.19 \\
Y II  &   3 &     &$+0.69$&$-1.52$&  0.11 &$+0.15$&  0.08 \\
Ba II &   3 &     &$+0.61$&$-1.57$&  0.13 &$+0.10$&  0.10 \\
La II &   1 &     &$-0.11$&$-1.21$&  0.16 &$+0.46$&  0.14 \\
Eu II &   2 &     &$-0.52$&$-1.04$&  0.11 &$+0.64$&  0.08 \\
Dy II &   1 &     &$+0.20$&$-0.91$&  0.17 &$+0.77$&  0.16 \\
\enddata
\tablecomments{The ``ul'' column is an upper limit flag.}
\end{deluxetable}

\newpage

\section*{Acknowledgements}

This paper includes data gathered with the 6.5~meter Magellan Telescopes located at Las Campanas Observatory, Chile.
We thank all the LCO staff for enabling astronomical observations during a global pandemic.
We thank Yuri Beletsky for assistance with part of these observations and Ting Li for helpful conversations.
This work benefited from the Gaia DR3 Chicago Sprint hosted by KICP.
A.P.J. thanks M.C.T. for her infinite patience.
R.P.N. acknowledges an Ashford Fellowship granted by Harvard University.
K.B. acknowledges support from the United States Department of Energy grant DE-SC0019323.
Y.S.T. acknowledges financial support from the Australian Research Council through DECRA Fellowship DE220101520. 

This work has made use of data from the European Space Agency (ESA) mission
{\it Gaia} (\url{https://www.cosmos.esa.int/gaia}), processed by the {\it Gaia}
Data Processing and Analysis Consortium (DPAC,
\url{https://www.cosmos.esa.int/web/gaia/dpac/consortium}). Funding for the DPAC
has been provided by national institutions, in particular the institutions
participating in the {\it Gaia} Multilateral Agreement.
This research has made use of NASA's Astrophysics Data System Bibliographic Services.

\section*{Data Availability}

The individual line measurements are provided as supplementary material. The reduced spectra can be obtained by reasonable request to the corresponding author.




\bsp	
\label{lastpage}
\end{document}